\documentclass{article}
\usepackage{graphicx}
\usepackage{url}
\usepackage{dcolumn}
\usepackage{upgreek}
\usepackage{authblk}

\newcommand{\etal}{et.al.}

\newcommand{\hn}{\ensuremath h_{90}}
\newcommand{\KJn}{\ensuremath K_{\mathrm{J-90}}}
\newcommand{\KJ}{\ensuremath K_{\mathrm{J}}}
\newcommand{\RKn}{\ensuremath R_{\mathrm{K-90}}}
\newcommand{\Itrim}{\ensuremath I_{\mathrm{Trim}}}
\newcommand{\Iskew}{\ensuremath I_{\mathrm{Skew}}}
\newcommand{\nV}{\ensuremath n_{\mathrm{V}}}
\newcommand{\Vth}{\ensuremath V_{\mathrm{th}}}
\newcommand{\VD}{\ensuremath V_{\mathrm{D}}}
\newcommand{\VU}{\ensuremath V_{\mathrm{U}}}
\newcommand{\vD}{\ensuremath v_{\mathrm{D}}}
\newcommand{\vU}{\ensuremath v_{\mathrm{U}}}
\newcommand{\rD}{\ensuremath q_{\mathrm{D}}}
\newcommand{\rU}{\ensuremath q_{\mathrm{U}}}

\newcommand{\mc}{\ensuremath m_{\mathrm{c}}}
\newcommand{\mn}{\ensuremath m}
\newcommand{\moffs}{\ensuremath m_{\mathrm{sp}}}
\newcommand{\zOn}{\ensuremath z_{\mathrm{On}}}
\newcommand{\zOff}{\ensuremath z_{\mathrm{Off}}}
\newcommand{\IOn}{\ensuremath I_{\mathrm{On}}}
\newcommand{\IOff}{\ensuremath I_{\mathrm{Off}}}
\newcommand{\deltaI}{\ensuremath \delta I}
\newcommand{\IA}{\ensuremath I_{\mathrm{A}}}
\newcommand{\Deltaz}{\ensuremath \Delta z}
\newcommand{\Barz}{\ensuremath \bar{z}}
\newcommand{\VOn}{\ensuremath V_{\mathrm{On}}}
\newcommand{\VOff}{\ensuremath V_{\mathrm{Off}}}
\newcommand{\nF}{\ensuremath n_{\mathrm{F}}}
\newcommand{\mSI}{\ensuremath \{m\}_{\mathrm{SI}}}

\newcommand{\niii}{\mbox{NIST-3}}
\newcommand{\mymu}{{\ensuremath \upmu}}

\newcommand{\hres}{\ensuremath 6.626\,069\,79(30)\times 10^{-34}}
\newcommand{\hrel}{\ensuremath 45\times 10^{-9}}
\newcommand{\hhn}{\ensuremath 141\times 10^{-9}}

\begin{document}
\title{Determination of the Planck constant using a watt balance with a superconducting magnet system at the National Institute of Standards and Technology}
\author[1]{S Schlamminger\thanks{stephan.schlamminger@nist.gov}}
\author[2]{D Haddad}
\author[2]{F Seifert}
\author[1]{L S Chao}
\author[1]{D B Newell}
\author[1]{R Liu}
\author[1]{R L Steiner}
\author[1]{J R Pratt}
\affil[1]{National Institute of Standards and Technology (NIST), 100 Bureau Drive Stop 8171, Gaithersburg, MD 20899, USA}
\affil[2]{Joint Quantum Institute, National Institute of Standards and Technology and University of Maryland, Gaithersburg, MD 20899, USA }
\date{}



\maketitle

\begin{small}
For the past two years, measurements have been performed with a watt balance at the National Institute of Standards and Technology (NIST) to determine the Planck constant. A detailed analysis of these measurements and their uncertainties has led to the value  $h=\hres\,$J\,s. The relative standard uncertainty is $\hrel$. This result is $\hhn$ fractionally higher than $h_{90}$. Here $h_{90}$ is the conventional value of the Planck constant given by $h_{90}\equiv 4 /( \KJn^2\RKn)$, where $\KJn$ and $\RKn$ denote the conventional values of the Josephson and von Klitzing constants, respectively.
\end{small}

\section{Introduction}
More than a quarter century ago, researchers at the National Institute of Standards and Technology (NIST) designed the first iteration of a watt balance for  the purpose of improving the realization of the SI unit of electrical current, the ampere~\cite{Olsen80a,Olsen80b}. A conceptually identical instrument was used in 2012-2013 to measure a precise value of the Planck constant, $h$. Using a device that was conceived to realize the ampere to measure the Planck constant was made possible by the discovery of the quantum Hall Effect (QHE) by von Klitzing in 1980~\cite{Klitzing80}. 

The existing watt balance is the third prototype~\cite{Olsen89,Steiner05,Steiner07} of such an instrument at NIST and is named \niii. One interesting feature of \niii\ stems from its legacy as a device to realize the ampere: the electro-magnetic part of the watt balance is entirely composed of coils and is void of ferromagnetic materials. This is in contrast to other existing watt balances, which use permanent magnets with ferromagnetic yokes to generate the required magnetic field. 

This design choice led to interesting advantages: The current in the superconductor can be varied, allowing operation of the apparatus with different magnetic fields. The large radial extent of the magnetic field keeps the spatial variation of the magnetic field smooth, and hence the field gradients small.

These advantages come with a price: The experiment has a relatively low duty cycle. The liquid helium dewar needs to be refilled three times a week. The apparatus is large in size to house the necessary coils and liquid helium. Because of its extent, the instrument requires frequent alignment. Since these maintenance tasks are performed during the day, data are obtained mostly at night, yielding a duty cycle slightly above 50\,\%.

Since \niii\ differs in size and magnetic field generation from other watt balances, we believe that the correlations between the result presented here and other results are limited to the correlations in the fundamental calibrations, mainly the mass calibration, that trace back to the Bureau International des Poids et Mesures (BIPM).

\section{Principle of measurement and main equations}
The NIST watt-balance experiment is an alternating series of measurements in two modes, velocity mode (V) and force mode (F), yielding a VFVFV...V series of measurements. 
In velocity mode, a circular coil is moved through a radial magnetic field. The induced voltage $V$ of the coil moving at velocity $v$ is 
\begin{equation}
V = Bl v,
\end{equation}
where $Bl$ is the flux integral, i.e., the integrated product of magnetic flux density and wire length of the coil. In addition to this induced voltage, spurious AC-signals, such as multiples of the power-line frequency, are induced in the moving coil. To suppress these spurious voltages, a stationary coil is connected in series opposition with the moving coil. In fact, two stationary coils are used, effectively forming one coil with a coupling to the environment identical to that of the moving coil in the center of the travel range.

\begin{figure}[htb]
\begin{center}
\includegraphics[width=3.25in]{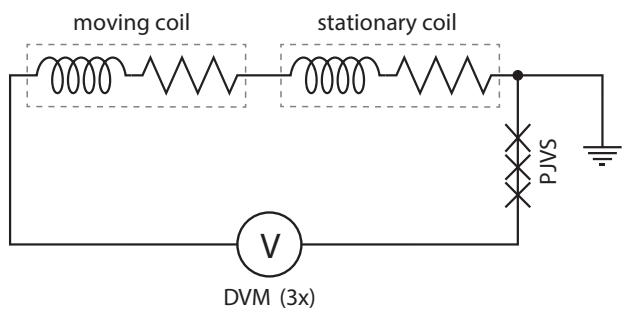}
\end{center}
\caption{The electrical circuit during velocity mode. The two stationary coils are drawn as one coil. The coils are depicted as an inductive and resistive element.}
\label{fig:cir:velo}
\end{figure}

The voltage induced by the motion of the coils needs to be measured with a relative standard uncertainty better than a part in $10^8$. This is accomplished with the help of a Programmable Josephson Voltage System (PJVS)~\cite{Benz97}. A PJVS generates a voltage, $V= f \nV/\KJn$, where $f$ is the microwave frequency, $\nV$ the number of Josephson junctions, and $\KJn$ the conventional Josephson constant~\cite{Taylor89}. The PJVS is connected in series with the moving and stationary coils, see figure~\ref{fig:cir:velo}.  A set of digital voltmeters (DVMs) is used to measure the  voltage difference between the coil and the PJVS. The voltmeter readings when the coil moves down (D) and up (U) are
\begin{eqnarray}
\VD &=&   \frac{ f \nV }{\KJn}+ Bl \vD+ \Vth,\\
\VU &=&  -\frac{ f \nV }{\KJn}+ Bl \vU + \Vth,
\end{eqnarray}
where $\vD$, $\vU$ denote the vertical velocity of the coil during the down and up motion, respectively. $\Vth$ denotes parasitic voltages, such as thermal voltages and zero points of the voltmeters. The $z$-axis points upward, hence $\vU>0$ and $\vD<0$. The velocities are measured simultaneously with the voltage measurement using three heterodyne interferometers. The polarity of the PJVS is switched between the down and up sweep by reversing the bias current to all junctions. All electrical measurements are performed in conventional electrical units~\cite{Taylor89}, hence $\KJn$ rather than $\KJ$ is used in the equations above.

From the known and measured quantities, two quotients are formed,
\begin{eqnarray}
\rD=\frac{ \VD - f \nV/\KJn}{\vD} &=& Bl + \frac{\Vth}{\vD},\\
\rU=\frac{ \VU + f \nV/\KJn}{\vU} &=& Bl + \frac{\Vth}{\vU}.
\end{eqnarray}
Ideally, $\VU$ and $\VD$ are close to zero ($\le 1$\,mV) to minimize gain errors in the voltmeters. These quotients are calculated along the 8\,cm long coil trajectory yielding  $\rD(z)$ and $\rU(z)$. The  flux integral can be inferred from the average of the two quotients
\begin{equation}
Bl(z) = \frac{1}{2} \Big( \rU(z) + \rD(z) \Big)  - \epsilon(z), \label{eq:bl}
\end{equation}
where $\epsilon(z)$ is a small error term that stems from an insufficient cancellation of the thermal voltages if $\vD \ne -\vU$. The relative magnitude of this term is well below $10^{-10}$, and it is ignored in the derivation below.

During force mode, two principal measurements, called mass off and mass on, are made in an alternating order. At the start of every force mode, a counter mass, $\mc$, is loaded on the tare side of the balance. This counter mass remains on the balance for the entire time of the force measurements. The mass pan is empty during the mass-off state and carries a mass, $m\approx 2\mc$, during the mass-on state. A current is passed through the moving coil resulting in a force $F=Bl I$. In each state, the current is controlled such that the balance maintains a chosen position, leading to  
\begin{eqnarray}
Bl(\zOff) \IOff &=& - \mc g - \moffs g \label{eq:bloff},\\
Bl(\zOn) \IOn -\mn g&=& - \mc g -\moffs g \label{eq:blon} .
\end{eqnarray}
The term $\moffs g$ denotes spurious forces on the balance that are present in both measurements, e.g., imbalance of the wheel and forces produced by the central pivot of the balance. The currents have different signs, i.e. $\IOff<0$ and $\IOn>0$. Subtracting the on state from the off state yields
\begin{equation}
\mn g = Bl(\zOn)\IOn-Bl(\zOff)\IOff. \label{eq:fm}
\end{equation}

\begin{figure}[htb]
\begin{center}
\includegraphics[width=3.25in]{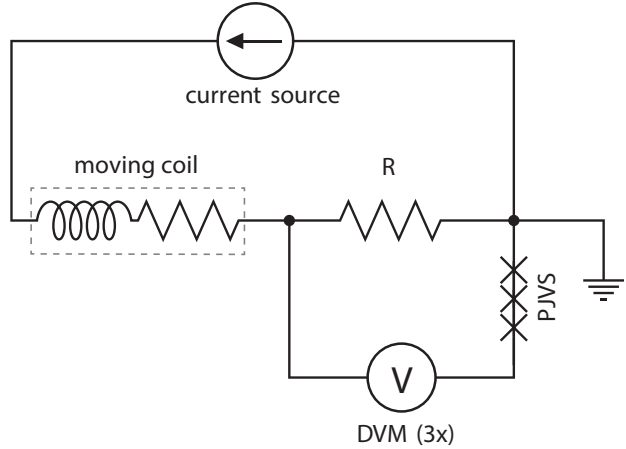}
\end{center}
\caption{The electrical circuit in force mode.}
\label{fig:cir:force}
\end{figure}

Note that even though the balance is controlled to the same position, the positions, $\zOn$ and $\zOff$ differ slightly  due to mechanical compliance in the coil support. During mass on, the coil is about $10\,$\mymu m  higher than during mass off. Although, as it is shown below, the change in coil position is insignificant for the result, it is taken into account in the data analysis. 

The circuit diagram for force mode is shown in figure~\ref{fig:cir:force}. The current flows through the moving coil and through a resistor $R$. The voltage drop across $R$ is compared to the PJVS. The readings of the DVMs for the two states are:
\begin{eqnarray}
\VOff&=&R \IOff  + \frac{f \nF}{\KJn} + \Vth,\\
\VOn&=&R \IOn    - \frac{f \nF}{\KJn} + \Vth.
\end{eqnarray}
The above equations can be solved for $\IOff$ and $\IOn$. However, it is advantageous to perform a coordinate transformation to obtain $\IA=\frac{1}{2}(\IOn -\IOff)$ and $\deltaI = \frac{1}{2}(\IOn+\IOff)$. $\IA$ is named the current amplitude and $\delta I$ the current asymmetry. $\IA\approx \IOn$ and $\deltaI\approx 0$. These two new variables can be computed from the measurements using 
\begin{eqnarray}
\IA &=& \frac{\displaystyle \Big( \frac{1}{2} (\VOn-\VOff) +\frac{\displaystyle f n_F}{\displaystyle \KJn}  \Big)}{\displaystyle R},\\
\deltaI  &=& \frac{\displaystyle \Big(\frac{1}{2}  (\VOn+\VOff) -  \Vth \Big)}{\displaystyle  R}.
\end{eqnarray}
 A different coordinate transformation is used for the coil position:
$\Barz = \frac{1}{2}(\zOn+\zOff)$ and $\Deltaz = \frac{1}{2}(\zOn-\zOff)$. 

Rewriting (\ref{eq:fm}) in these new variables and expanding to second order yields
\begin{eqnarray}
\mn g &\approx& 2 Bl(\Barz) \IA + \left. 2 \Deltaz \deltaI\frac{\mbox{d}(Bl)}{\mbox{d}z}\right|_{z=\Barz} \nonumber \\
& & +  (\Deltaz)^2 \IA  \left. \frac{\mbox{d}^2 (Bl)}{\mbox{d}z^2}\right|_{z=\Barz}. \label{eq:fm2}
\end{eqnarray}
The second and third term on the right side of the equation are small and can be neglected. Using $Bl$  from velocity mode, the mass can be obtained:
\begin{equation}
\mn \approx \frac{ (\rU+\rD) \IA }{g} \label{eq:m90}.
\end{equation}
Since all measurements on the right side of the equations are carried out in conventional units, we  obtain the value of $m$ in conventional units. Note that the units of time and length are identical in conventional and SI units.

The ratio of $h$ to $\hn$ is given by the ratio of the numerical value of the mass in SI units, $\{m\}_\mathrm{SI}$, to the value of the mass in conventional units, ${\{m\}_\mathrm{90}}$. The former is obtained by the mass group at NIST and  the latter by (\ref{eq:m90}). Thus, the SI Planck constant is given by
\begin{equation}
h = \frac{\{m\}_\mathrm{SI}}{\{m\}_\mathrm{90} } h_{90}, \label{eq:h90}
\end{equation}
where $\{m\}_\mathrm{SI}$ and $\{m\}_\mathrm{90}$ denote the numerical values of the quantity $m$ in SI and conventional units, respectively. 
The conventional value of the Planck constant is given by~\cite{Taylor89}
\begin{equation}
h_{90} \equiv \frac{4}{\KJn^2 \RKn}  =  6.626\,068\, 854\,\ldots\;\times\; 10^{-34}\,\mbox{J\,s~}.
\end{equation}

\section{Description of the apparatus, NIST-3}
\begin{figure}[htb]
\centering
\includegraphics[width=3.25in]{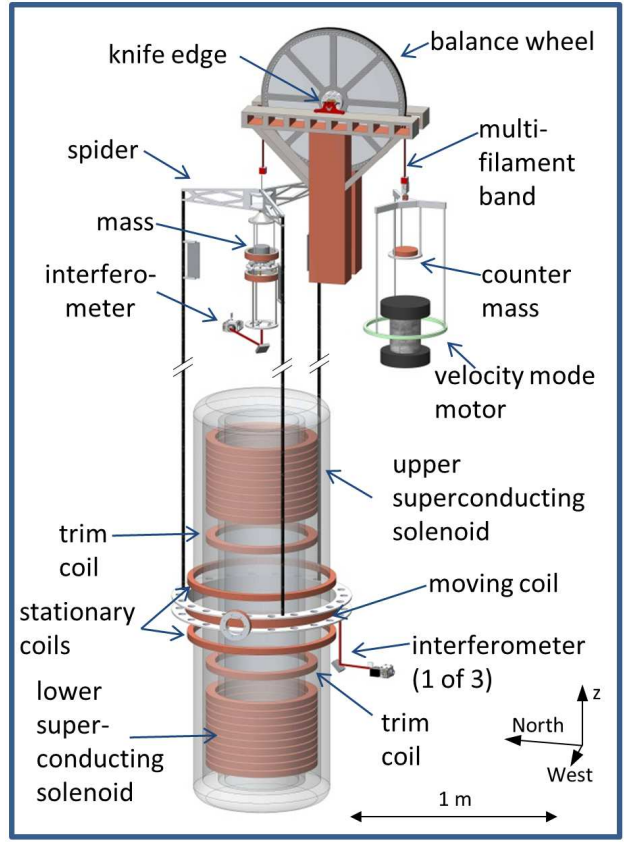}
\caption{Drawing of the principal components of \niii. For clarity, the vacuum envelope around the top part of the apparatus and the moving coil has been omitted. In reality, there is more space between the coils and the spider, indicated by breaks in the rods connecting the spider to the moving coil.}
\label{fig:NIST3:schema}
\end{figure}

Figure~\ref{fig:NIST3:schema} shows a schematic drawing of the \niii\ watt balance. Details of the apparatus can be found in~\cite{Steiner05b}. The apparatus spans two stories which are separated by a false floor not shown in the drawing. The balance is located in the top room and the magnet system in the lower room. 

The balance itself is a wheel with 30.5\,cm radius and 2.54\,cm thickness. The wheel is made of aluminum and has a stainless steel ring around the circumference to provide a hardened surface for two multi-filament bands that connect to two loads on either side of the wheel. The center of the wheel is supported by a 7.62\,cm long knife edge made from tungsten carbide resting on a flat made from the same material. Both the flat and the knife edge are coated with diamond like carbon to avoid sticking. The knife edge and flat provide a one dimensional frictionless pivot point and allow the wheel to rotate around its center by $\pm10^\circ$. 

The multi-filament band on the north side of the wheel connects to a part called the spider. The moving coil is suspended from the spider by  three 4.2\,m  long rods. The mass pan is located below the spider and is connected to the spider via a copper-beryllium flexure cube. The spider itself is connected using a similar flexure to the multi-filament band. These flexures allow for nearly decoupled pivoting of the mass pan and the spider about the north-south (NS) and east-west (EW) axes. The mass pan is nested between a set of small coils that can be used to investigate magnetic properties of the mass. Directly attached to the spider and below the mass pan is a corner-cube retroreflector  that reflects a laser beam of a heterodyne interferometer. The readout of this interferometer is used as the control input for the balance servo during force mode.

Three additional interferometers monitor the position of the moving coil. Their measurement beams are reflected from three hollow retroreflectors mounted on the coil spaced $120^\circ$ apart. The readouts from the detectors are combined to calculate the velocity of the mass center of the coil during velocity mode. 

Twenty two superconducting coils are used to produce the radial magnetic flux density at the moving coil~\cite{Olsen80a,Chen82}.  The coil system is up-down symmetric about the moving coil, when the balance is controlled to the weighing  position. Twenty coils form two long solenoids and two coils are used as trim coils.
A current of  approximately $5\,$A flows through  all 22 coils but with opposite direction through the lower 11 coils. The current is not persistent but is stabilized by a control circuit that controls a voltage source based on the measured difference between the voltage drop caused by this current across a 0.2\,$\Omega$ resistor and a Zener stabilized voltage source. Two additional current sources are used: a trim current, $\Itrim\approx 59$\,mA, flows only through the trim coils and a skew current, $\Iskew\approx 13$\,mA, flows only through the upper solenoid. The trim and skew currents are in addition to the 5\,A. Both trim and skew currents are used to correct for physical imperfections of the superconductor. The trim current is adjusted to bring the radial flux density as close as possible to $B_\mathrm{r}\propto 1/r$. The purpose of the skew current is to make the measured field up-down symmetric.

The moving coil has 2478 turns, a mean diameter of 0.713\,m and a height of 0.043\,m. The inductance of the coil is $L=9.14$\,H and the DC resistance is 752\,$\Omega$. In normal operation, the flux integral is 492\,T\,m.

The balance and the moving coil are enclosed in two fiberglass vacuum vessels that are connected by three pipes around the rods connecting the moving coil to the spider. The vacuum hardware is omitted in the drawing. The vacuum pressure is typically 0.5\,Pa. The vacuum is necessary to reduce measurement biases due to buoyancy and index of refraction.

The platinum-iridium prototype No.85, referred to as K85, is the mass standard used in the experiment. This prototype was purchased in 2003 from the BIPM. In the beginning of 2012, another calibration and an air-vacuum transfer study was performed at the BIPM. Shortly before the calibration, the prototype was cleaned twice using the BIPM method. During the course of the experiment, the prototype was measured five times by the mass  group at NIST. The mass remained stable within the Type A uncertainty at mass value of $1\,\mbox{kg}-738.3\,\mymu$g with a relative standard uncertainty of $ 9.6\times 10^{-9}$.

Two different resistors are used in the experiment. Both are wire wound resistors made from Evanohm\footnote{Evanohm is a registered trademark of CSR Holdings, Inc.}$^{,}$\footnote{Certain commercial equipment, instruments, or materials are identified in this paper in order to specify the experimental procedure adequately. Such identification is not intended to imply recommendation or endorsement by the National Institute of Standards and Technology, nor is it intended to imply that the materials or equipment identified are necessarily the best available for the purpose}. The resistors are kept in an oil bath at 25\,$^\circ$C. The resistors were calibrated before and after usage in the watt balance against a Quantum Hall Effect resistance standard in the NIST resistance laboratory~\cite{Delahaye00}. The calibration was performed in conventional units.

\section{A brief summary of changes made since January 2012}

NIST-3 produced stable values for $h$ from October 2004 to March 2010 when the value suddenly increased to a second value (see section~9). In order to investigate the reason for this or to find which of the two values is correct, the 2012-2013 measurement was initiated with the goal to produce an independent value of the Planck constant. This measurement was conducted blindly by keeping the exact value of the mass $\mSI$ hidden from the experimenters. The idea of this blind measurement was to eliminate experimenters' bias toward previously measured Planck values.

The mass group added a constant, relative bias between  $-500\times 10^{-9}$ and $+500\times 10^{-9}$  to all mass calibrations communicated to the watt-balance team. However, the exact value of this bias was intentionally withheld from the experimenters. The bias was revealed during a public talk in June 2013.

Before the measurements described here were taken, the watt balance was closely inspected and several changes were made. The changes include:

\begin{itemize}
\item The power conditions were improved. It was found that the power-line voltage was below 110~V. The power-line voltage was restored by changing settings in the un-interruptible power supply (UPS) that conditions the power for the watt balance.
\item Power-line filters were installed at the three penetrations into the shielded room which houses the watt balance. The insertion loss of each  filter is at least 100\,dB in a frequency range from 14\,kHz to 10\,GHz.
\item The grounding system was changed into a star topology.
\item Questionable contacts in the grounding and shielding were improved. 
\item The electrical wiring was re-organized. Obsolete wires and instruments were removed.
\item The three DVMs used in the experiment were replaced with new ones.
\item The PJVS was upgraded: The communication between the PJVS current source and the control computer was changed from the parallel port to a dedicated PCI digital input/output card. A new microwave amplifier was installed.
\item A new current source~\cite{Haddad12} was installed. The current source was replaced again during the summer of 2013 with an upgraded version.
\item The knife edge and the flat were replaced.
\item The mass, K85, was sent to the BIPM for cleaning and re-calibration. 
\end{itemize}

Most of the changes were made to the electrical system, since the shift in value in 2010 was coincident with modifications of the electrical measurement system. 

\section{Alignment, measurement, and samples of data}

To begin a watt balance measurement, ten alignment steps are required, usually taking about one week to complete. From the beginning, the magnetic field is turned off. 
\begin{enumerate}
\renewcommand*\labelenumi{(\theenumi)}

\item The balance mechanics are adjusted to minimize the horizontal and angular velocities of the moving coil during up and down motion, see~\cite{Gillespie97}.

\item  The axis of the superconductor is aligned to be vertical by measuring the mutual inductance between the superconductor and a homemade coil assembly mounted on a leveling platform~\cite{Williams72}. The angle of the field can be adjusted by tilting the dewar containing the superconducting coil. 

\item  The electrical center of the moving coil is made coincident with the center of the magnetic field generated by the superconductor. The measurement for this alignment step is again a mutual inductance measurement. This time an alignment instrument with three individual coils is used. The electrical center of the coil can be moved by translating the watt balance wheel. After this step, the field is turned on.

\item  The tilt of the moving coil is changed such that the coil does not experience any horizontal forces when current is applied to the coil, indicating that the magnetic axes of both the superconducting coil assembly and the moving coil are vertical. However, this condition can also be met with both coils inclined in opposite directions. Since step (2) aligns the axis of the superconducting magnet to vertical, step (4) ensures that the magnetic axis of the moving coil is also vertical. 

\item  The electrical center of the moving coil is rechecked by observing a tilt when the coil is energized with $\pm$10\,mA. This procedure has more resolution than step (3).

\item  The positions of the three interferometers relative to the center of mass of the coil are determined by exciting tilt motion of the moving coil. A least squares adjustment of the readings of the three interferometers to a model~\cite{Haddad10} allows us to construct a matrix that converts the interferometer readings into the vertical position of the center of mass and the tilt around two axes. 

\item  The suspension point of the mass pan is adjusted such that the coil tilt is minimal when the mass is placed on the mass pan. This is performed without current in the coil. After that, the procedure is repeated with the balance in feedback, i.e., current in the coil.

\item  The mass lift is adjusted such that disturbance of the balance during mass transfer is minimal. 

\item  The three measurement beams of the lower interferometer are aligned to the local vertical direction.

\item The whole balance is moved such that the change in tilt of the moving coil between the mass-off and mass-on state is minimal. 

\end{enumerate}

The data presented here is organized in campaigns, each one lasting between three days and three weeks. Each campaign consists of several runs. A run was typically started at 6~pm and ended the next morning at 9~am.  A typical run during the week consists of 14 velocity-mode measurements interleaved with 13 force mode measurements. On weekends, longer runs were made.

The alignments in steps (1) to (3) of the above procedure were performed at the beginning of the experiment and two more times, about every 6 months. Before each campaign, the procedures in steps (4) through (10) were executed in an iterative manner to converge on a final alignment.  During a campaign, step (9) was performed three times a week and step (10) every day before the start of a measurement run. Repeating step (10) was necessary to compensate mechanical relaxation and changes in the system. 

\begin{figure}[htb]
\centering
\includegraphics[width=3.25in]{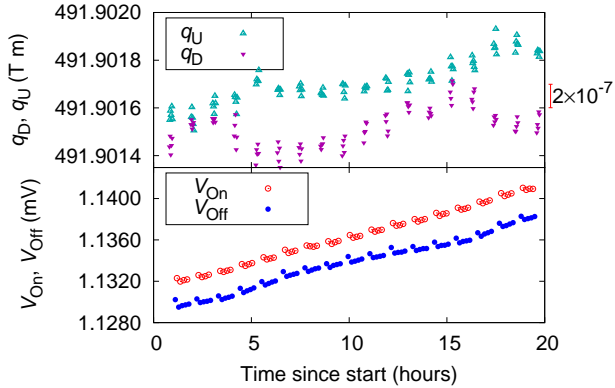}
\caption{Data of a typical run. The top plot shows the measured voltage-velocity quotient. In each group, ten measurements are performed, five with the coil moving up, five with the coil moving down. The bottom plot shows the readings of the voltmeter during the mass-off (solid circles) and mass-on (open circles) measurements.}
\label{fig:measurement}
\end{figure}

Figure~\ref{fig:measurement} shows a 20\,h run as an example. The top panel shows the measured quotients during  velocity mode.  The mean value is the estimate of the flux integral, $Bl$. Note that for the real data analysis, the $Bl$ is evaluated at $\zOn$ and $\zOff$. For this graph, however, the field profile is evaluated at a nominal position. The flux integral changes slowly during the course of the experiment for various reasons. These changes are slow so that the flux integral between two velocity modes can be approximated by linear interpolation between two adjacent measurements. The difference between the up and the down group is caused by the thermal voltage, $\Vth$. The changes in $\Vth$ are correlated with changes in the room temperature.

The bottom plot in figure~\ref{fig:measurement} shows the measured voltages $\VOn$ and $\VOff$ during force mode. The change in force on the balance between the two states is given by the difference of the two voltages. Note that the absolute values of the voltages are small, $\approx1\,$mV, and the difference between the two states is $\approx 2\,\mymu$V, thus minimizing gain errors. It can be seen that the first measurement in each force-mode group deviates from the long term drift. This is caused by a relaxation process in the knife edge. In order to account for this, the first points are discarded. There is a small effect on the remaining points. We found that the result is influenced by the motion of the balance, during the switch from velocity mode to force mode. If the balance is perfectly balanced, $\moffs g=0$, no bias was found. We apply a small correction, see table~\ref{tab:Corrections}, to account for this dynamic knife edge hysteresis. More detail on the knife edge hysteresis can be found in section 7.

\begin{figure}[htb]
\centering
\includegraphics[width=3.25in]{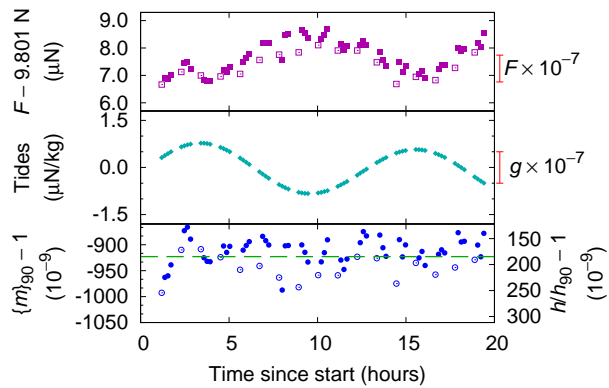}
\caption{Force measurements are shown in the top plot. This data is obtained from the difference in $\VOn$ and $\VOff$, the measured flux integral and the calibration value of $R$. In the middle plot,  the results of a tidal calculation~\cite{Wenzel96,Newell13} are shown. The calculation is performed using the time of each force measurement. The calculated values of $\{\mn\}_{90}-1$ using (\ref{eq:m90}) are shown in the bottom plot. The first points in each force mode group (open symbols) are discarded in the final analysis.}
\label{fig:measurement2}
\end{figure}

The measurements from velocity mode and force mode are combined to calculate the force produced by the mass using (\ref{eq:fm2}). The result can be seen in the top panel of figure~\ref{fig:measurement2}. During the measurement, the local acceleration changes due to tides~\cite{Wenzel96,Newell13}, air pressure variations in the atmosphere, and polar motion. Correction values for these three terms are calculated and added to the data.  The middle panel of figure~\ref{fig:measurement2} shows the largest of these three, the tidal correction.  The tidal correction is calculated using the software package QuickTide Pro from Micro-g LaCoste, Incorporated. To be confident in the tidal correction, we measured the absolute value of $g$ for six weeks in the vicinity of the apparatus using an absolute gravimeter, FG5-204 made by Micro-g LaCoste, Incorporated. We found no significant difference between the tides model and the measurements.
After the variations in $g$ have been applied and the nominal value of the gravitational acceleration $g=9.801\,016\,681\,\mbox{m}\,\mbox{s}^{-2}$ at the center of the mass is divided out, values of $\{\mn\}_{90}$ can be obtained; see the lower panel of figure~\ref{fig:measurement}. These values are inversely proportional to the Planck constant; see (\ref{eq:h90}). The right axis of the lower plot shows the values of $h/\hn-1$, using $\mSI-1=-738 \times 10^{-9}$.

\section{Measurements of the Planck values}

A total of six measurement campaigns were performed using the prototype K85 during 2012 and 2013. The first four campaigns were performed blindly, i.e., the experimenters did not know the precise value of the mass, $\mSI$. After the true value of $\mSI$ was revealed, two additional campaigns were performed. Between the first four campaigns and the last two campaigns, the superconducting magnet system was warmed to room temperature and cooled back down to liquid helium temperatures. During this measurement break, several other changes were implemented: (1) The receivers for the lower three interferometers were replaced with fiber-coupled optical receivers. (2) The current sources for the moving coil in force mode and for the motor driving the balance in velocity mode  were replaced. (3) Some software parameters for the mass exchange were improved. 

\begin{figure}[htb]
\centering
\includegraphics[width=3.25in]{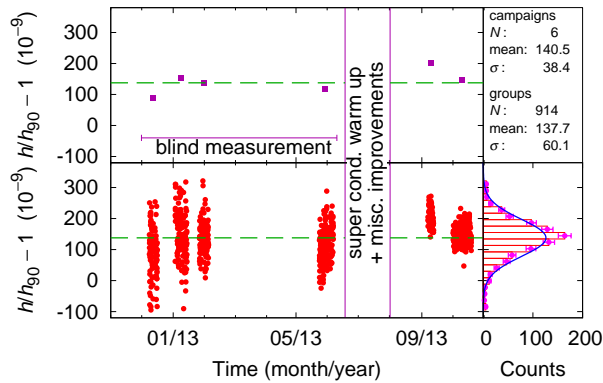}
\caption{Results of six measurement campaigns. The top plot shows the average number obtained for each campaign. In the bottom plot, a dot marks a result from each force mode measurement. On the top right panel, the mean and the standard deviation of the data point are given. The lower right panel shows a histogram of force mode data. The error bars in the histogram denote the 1-sigma statistical uncertainty assuming a Poisson distribution of the frequency in each bin.}
\label{fig:result}
\end{figure}

The results of the six measurement campaigns are shown in figure~\ref{fig:result}. The lower plot shows one point for each force mode measurement. On the top, the mean value of each campaign is shown. The unweighted average of the six results agrees well with the average of all force mode data. 

The data of the six campaigns show more variance than is expected from the standard deviation of the data points of the force-mode groups. This hints at a variation of the system from campaign to campaign. Unfortunately, we were not able to correlate the change in the $h$ value with any other observation. Hence, we include the effect as a Type A uncertainty in the measurement. We use the standard deviation of the 6 campaigns and divide by $\sqrt{6}$ to obtain the Type A uncertainty of the measurement of $15.7\times 10^{-9}$.

\section{Investigations of systematic effects}
The data contained in the six measurement campaigns were taken in about 10 weeks. Ten times more time was spent on aligning, maintaining, and improving the instrument and investigating systematic effects, e.g., ~\cite{Tang12}. Since describing all systematic measurements that were performed is beyond the scope of this article, we focus on the investigations of the two most controversial aspects of this experiment: the superconducting magnet and the knife edge.

The historic results (prior to 2009) from \niii\ differ from  a recent result~\cite{Steele12} produced  by the NRC (National Research Council, Canada) watt balance, which was formerly owned by NPL (National Physical Laboratory, UK)~\cite{Robinson12}. A logical starting point for understanding this difference is to investigate the main difference between the two watt balances, which is the source of the magnetic field. In the NRC watt balance, the magnetic flux is generated by Samarium-Cobalt magnets, while \niii\ uses a superconducting coil system to generate the flux through the moving coil.

An important assumption in most watt balance experiments is that the flux integral in  velocity mode is the same as in force mode. However, this is not obvious, since the moving coil carries current in force mode, but not in velocity mode. Any dependence of the magnetic field generation on this current would invalidate this assumption. An accepted parameterization~\cite{Robinson07} of the flux integral as a function of current in the moving coil is
\begin{equation}
Bl(I) = Bl(0)\big(1+\alpha I + \beta I^2 \big).
\end{equation}
Here $Bl(0)$ is the flux integral when no current is flowing in the moving coil, as is the case in velocity mode.
With this parameterization, (\ref{eq:fm}) can be written as,
\begin{equation}
\mn g = Bl(\IOn)\IOn-Bl(\IOff)\IOff,
\end{equation}
where the dependence of $z$ has been omitted for simplicity. Applying the coordinate transformation of the current yields
\begin{equation}
\mn g = 2 Bl(0) \IA \Big(1 + 2 \alpha \deltaI  + \beta \IA^2  + 3 \beta (\deltaI)^2 \Big).
\end{equation}

For all six campaigns, $\IA$ and $\deltaI$ are on average $10.0\,$mA and $-3.1\,\mymu$A, respectively.
The value of $\alpha$ was measured by adding 1\,g to the counter mass and comparing results measured with this additional load to normal measurements. This procedure changes $\deltaI$ by 20\,$\mymu$A. The measured values of $h$ for these two states was identical within uncertainties. 

The terms $\IA^2 \beta$ and $3 (\delta I)^2 \beta$ are more difficult to estimate. Since $\IA^2$ is about 300\,000 larger than $3 (\delta I)^2$, the latter can be ignored.

One suggestion for assessing $\beta \IA^2$ is to perform asymmetric weighings, i.e., $\IOn=20$\,mA and $\IOff=0\,$mA. However, such a test would destroy the symmetry of the watt balance experiment and its outcome is a convolution of several effects that will not cancel as well as they do in the symmetric case, e.g., coil heating and linearity of the DVM.

We took advantage of the fact that $Bl$ can be changed in \niii\ by simply changing the current in the superconducting coil.  These experiments were performed using a half kilogram stainless steel mass and not a one kilogram PtIr mass. Three measurements were performed in an ABA fashion. During the A and B states, the flux integral was 492\,T\,m and 196\,T\,m, respectively. $\IA$ is inversely proportional to $Bl$, since the same mass is balanced, i.e., more current in the moving coil is required if the magnetic field is weaker. For the states A and B, the coil currents were $\IA=5\,$mA and $\IA=12.5\,$mA. The difference in the squared value of $\IA$ is $131\,\mbox{mA}^2$.  A difference in $(h/h_{90}-1)$ of $(7.1\pm 24.5) \times 10^{-9}$ was obtained for the two states, with the $h$ being higher with $\IA=5\,$mA. 

The resolution of this experiment is limited by the type A uncertainty, and we cannot make a stronger statement than $|\IA^2 \beta|<25.5\times 10^{-9}$  with 68.8\,\% probability. Scaling this to the usual $\IA^2=100\,\mbox{mA}^2$, we attribute a relative uncertainty of $19.4\times 10^{-9}$ to the magnetic field.

Up to this point, the knife edge was treated as an ideal pivot point. In particular, it was assumed that $\moffs g$ remains identical between the mass-off and mass-on measurement in (\ref{eq:bloff}) and (\ref{eq:blon}). Unfortunately this assumption is not true. 

A systematic error can be introduced in the measurement in the following way: 
the balance wheel rotates by a small amount during mass exchange, i.e., the process that adds or removes the mass to or from the mass pan. The wheel rotation angle varies between $66\,\mymu$rad and $330\,\mymu$rad. This motion alters  the knife-edge flat interface and leads to a change in the forces on the balance, i.e., $\moffs g$ changes. If the motion of the balance is the same during mass-on and mass-off exchange, these offset changes are the same and cancel in the difference. However, the motion is, in general, different between the two types of mass exchanges. In order to further minimize the  effect of the mass exchanges, the balance is forced into a damped sinusoidal motion with initial amplitude of at least $260\,\mymu$rad after the exchange. The intent of this procedure is to erase the force that was imprinted to the knife edge during mass exchange. Note that the excursion of the wheel during mass exchange varied throughout the experiment, depending on alignment, mechanical compliance, and software settings. The amplitude of the damped sinusoid was adjusted accordingly. For most of the mass exchanges the excursions of wheel were less than the initial amplitude of the damped sine. In only about about 5\,\% of the cases the excursion was larger. Discarding this data would increase the measured value of $h$ fractionally by $1.7\times 10^{-9}$. We decided not to discard the data and it is included in the result presented below. These concerns with the so-called knife-edge hysteresis and the importance of the erasing procedure were recognized as early as 1991~\cite{Olsen91}. A summary of a systematic investigation into this problem is given in~\cite{Schwarz01}. 

Despite these precautions, a systematic error can remain if two conditions are met: (1) The erasing procedure does not significantly reduce the offset force. 
(2) The spurious wheel rotation is not the same when adding the mass as when removing the mass. Both conditions were investigated but were not found to be stable over time. It was found that the erasing procedure reduces the offset force by a factor of two. However, a significant relative bias can remain. The difference in balance motion between the mass-on and mass-off exchange was statistically analyzed. It was found that for every $10\,\mymu$rad of motion the value of $h$ changed by $3.6\times 10^{-9}$. Since the average difference in balance motion for the two type of exchanges is $33\,\mymu$rad a bias of $12\times 10^{-9}$ is introduced to the measurement. Due to the lack of consistency, no correction is applied to the data. Instead $12\times 10^{-9}$ is treated as an uncertainty in the category of ``balance mechanics''. Other contributions to this category include forces that arise from tilt changes of the coil between mass on and mass off, relaxation of the knife edge after velocity mode, and dependence of $h$ on the balance servo position.

\section{Final results and uncertainties}
To obtain the final result of our measurement, several corrections are applied to the data. Table~\ref{tab:Corrections} shows the corrections. The numbers shown in the table are the mean values of the corrections averaged over the data taking period. However, all corrections are applied individually to each data point gathered in force mode. All corrections sum up to the fractional amount of $-5.3\times 10^{-9}$, i.e., without applying the corrections, the result of $h$ would be fractionally $5.3\times 10^{-9}$ higher. 
The uncertainties of the corrections are included in the error budget.

\begin{table}[htb]
\begin{center}
\begin{tabular}{lD{.}{.}{2}}
Source & \multicolumn{1}{r} {Fractional} \\
& \multicolumn{1}{r}{correction} \\ 
& \multicolumn{1}{r}{to $h\;(10^{-9})$} \\ 
\\
Polar motion on $g$									&  +6.4 \\
Dynamic knife edge hysteresis						&  -6.3 \\
Alignment											&  -3.3	\\
Water desorption on mass							&  -3.1	\\
Diffraction of interferometer beams 	&  +2.8 \\
Air pressure variations on $g$						&  -2.1	\\
Verticality	of the interferometer beams 			&  +2.1	\\
Refractive index of residual air					&  -1.2	\\
Tidal variation of $g$								&  -0.8	\\
PJVS leakage										&  +0.4	\\
Buoyancy on the mass by residual air				&  -0.2 \\
Magnetic forces on K85							&  +0.1 \\
DVM gain correction							&   0.0 \\
\\
Total			& -5.3 \\
\end{tabular}
\end{center}
\caption{Average values of the corrections used for the six campaigns  discussed in this article. The sign of each correction is defined as follows: the corrections need to be added, including the sign, to the $h$ values in figure~\ref{fig:measurement2} to obtain the final values, shown in figure~\ref{fig:result}. Note that the tidal correction is an exception, since it is already added into the results shown in figure~\ref{fig:measurement2}.}
\label{tab:Corrections}
\end{table}

The error budget of the measurement is summarized in table~\ref{tab:uncert-budget}. The uncertainties are organized in eight categories.  The largest five items have combined Type A and B contributions ranging from $15.7\times 10^{-9}$ to $21.6\times 10^{-9}$. Hence, in order to significantly reduce the uncertainty of the experiment, work on all five categories is needed.

\begin{table}[htb]
\begin{center}
\begin{tabular}{lD{.}{.}{1}D{.}{.}{1}}
Source & \multicolumn{1}{c}{Type A} & \multicolumn{1}{c}{Type B} \\
& \multicolumn{1}{c}{$(10^{-9})$}  & \multicolumn{1}{c}{$(10^{-9})$}  \\
\\
Balance mechanics			& 5.0	&20.9	\\
Alignment					& 0.0	&20		\\
Magnetic field				& 19.4	&0	\\
Electrical 					&1.5	& 16.1\\
Statistical					&15.7	&0.0	\\
Velocity					& 0.0	&10.6		\\
Mass metrology				& 0.8	&9.6	\\
Local acceleration, $g$		& 3.8	&6.0		\\
\\
\\
Total	& 25.7& 36.6 \\
\\
\\
Type A \& B combined & \multicolumn{2}{c}{44.7}

\end{tabular}
\end{center}
\caption{Sources of uncertainty in this measurement of $h$. All entries are relative standard uncertainties $(k=1)$.}
\label{tab:uncert-budget}
\end{table}

\begin{figure}[htb]
\centering
\includegraphics[width=3.25in]{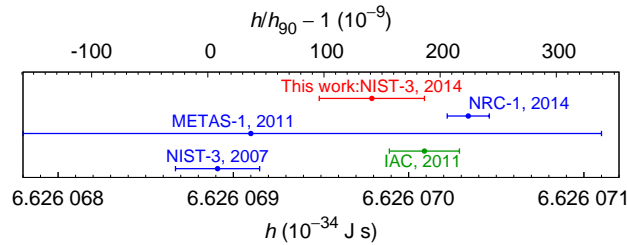}
\caption{The result of this work in comparison with other recent determinations of the Planck constant. The labels NRC-1, METAS-1, IAC, and \niii\ refer to results published in \cite{Sanchez14,Eichenberger11,Andreas11}, and \cite{Steiner07}. The dates indicate the year of publication. To convert the result of the IAC, which is a value of the Avogadro constant, to a value of the Planck constant, the CODATA adjusted value~\cite{CODATA10} of the molar Planck constant, $N_{\mathrm{A}}h=3.990\,312\,7176\,$J\,s\,mol$^{-1}$ was used. According to~\cite{CODATA10}, the relative standard uncertainty of the molar Planck constant is $7\times 10^{-10}$ and insignificant for the comparison shown in this figure.}
\label{fig:result_comp}
\end{figure}

Combining the results of the six measurement campaigns  with a careful evaluation of the uncertainties leads to 
\begin{equation}
\frac{h}{\hn} - 1 = 141(45) \times 10^{-9}.
\end{equation}
Hence, the value of the Planck constant is
\begin{equation}
h = \hres\,\mbox{J\,s}.
\end{equation}
The number in the parentheses denotes the one-sigma uncertainty in the last two digits.
Figure~\ref{fig:result_comp} shows this result in comparison to other recently published values of the Planck constant. There is good agreement between this result and the result from the International Avogadro collaboration (IAC). Our result is less than two combined standard deviations below the new value determined by experimenters at the National Research Council of Canada(NRC) using a watt balance~\cite{Sanchez14}.

\section{Notes on historical watt values from \niii}

\label{sec:hist}
\begin{figure}[htb]
\centering
\includegraphics[width=3.25in]{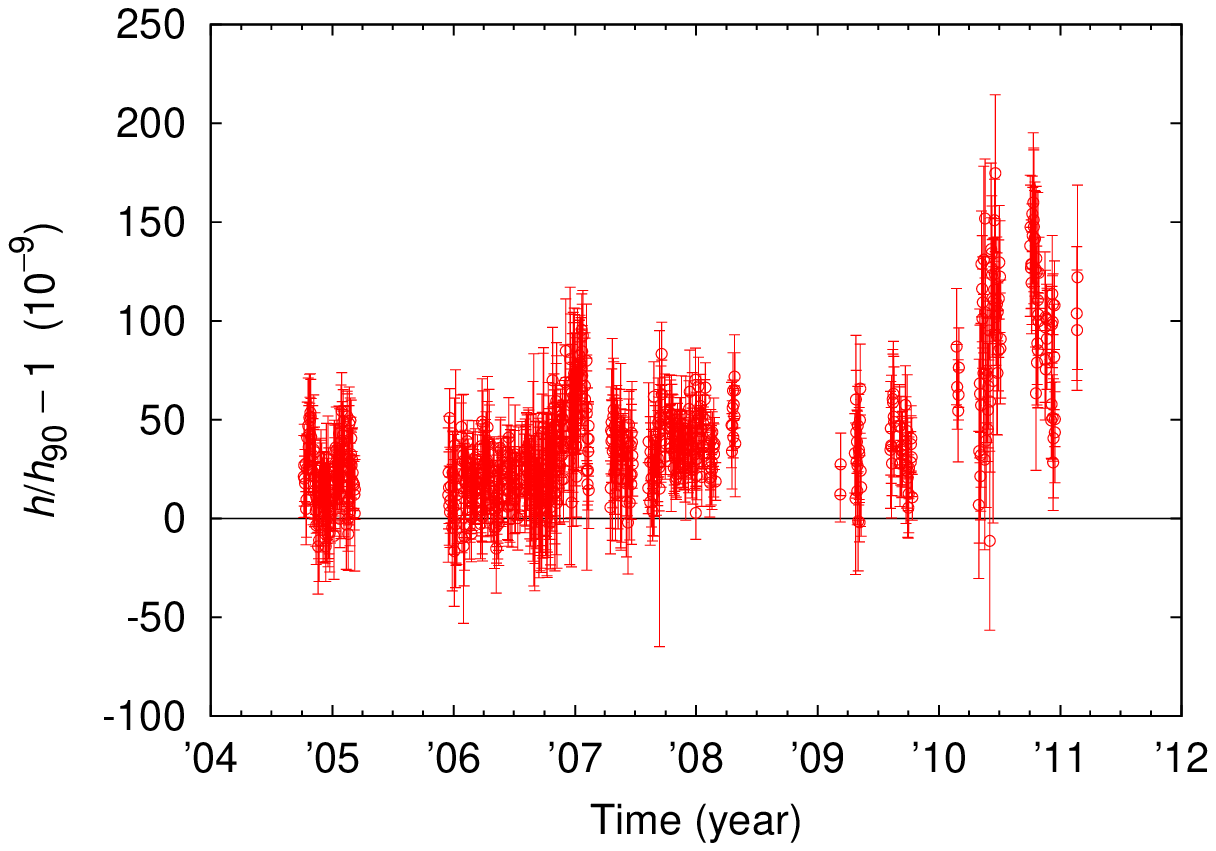}
\caption{Results obtained from October 2004 to February 2011. The data shown here were all taken using K85 as the mass standard. Error bars denote 1-sigma standard deviations of each daily run. In March 2010, a shift in value occurred coincident with an increase in noise.}
\label{fig:old:data}
\end{figure}

The construction of \niii\ started in 1999, shortly after the final result of NIST-2 was published~\cite{Williams98}. Initial measurements with \niii\ were performed as early as 2003. By 2004, the apparatus produced repeatable results using a gold mass standard, that was later replaced with the platinum-iridium prototype K85. Figure~\ref{fig:old:data} shows the results from all measurements with K85 from October 2004 to February 2011.

The mean value of the data taken before March 2010 is $h/\hn-1=29\times 10^{-9}$, consistent with the value published in 2005~\cite{Steiner05b} and 2007~\cite{Steiner07}, i.e., $h/\hn-1=(24\pm 52)\times 10^{-9}$ and $h/\hn-1=(8\pm 33)\times 10^{-9}$, respectively. A change seemed to have occurred during the time between March 2010 and May 2010. The mean value of the data after the break is $h/\hn-1=97\times 10^{-9}$. Coincidental with the shift in value is an increase in noise: the standard deviation of the data before and after March 2010 is $19\times 10^{-9}$ and $37\times 10^{-9}$, respectively. 

In March 2010, changes to the electrical measurement systems were made. While we cannot attribute the shift in the data to these modifications, they were the most significant changes to the system during that time.

Immediately after the shift had occurred, many subsystems were thoroughly investigated and improvements were made:

\begin{itemize}
\item The PJVS system was overhauled and the chip replaced.
\item The electrical grounding scheme was altered.
\item The isolation resistances to ground of the PJVS and of the current source were increased.
\item The oil in the resistor bath was changed.
\item Brass pieces on the balance that became magnetized in the field of the superconducting magnet were replaced with aluminum pieces.
\item Alignment procedures were improved.
\item The knife edge and flat were replaced.
\item Various mass exchanging and knife-edge hysteresis erasing procedures were tested.
\end{itemize}

Despite these improvements, the experiment continued to produce higher results with more variance in the data. None of the investigations found a plausible reason for the shift, nor any indication of which one of the two values is correct.

After the measurements shown in figure~\ref{fig:old:data} were completed, K85 was sent to the BIPM for recalibration. It was found that the mass of K85 had increased from a previous calibration in July 2010 by 40\,$\mymu$g. This mass increase is discrepant with the rate of change of $\approx 5\,\mymu$g per year as observed by NIST over the previous years. Note that the data in figure~\ref{fig:old:data} has not been updated to reflect the new BIPM calibration point, since it is difficult to know when this change occurred. We note that adding the $40\times 10^{-9}$  relative mass increase of K85 to the measured value of $h$ after March 2010 increases the value of $h/\hn-1$ to $137\times 10^{-9}$, which is in close agreement to the value of the 2012-2013 determination.

\section{Conclusion}
In 2012-2013, measurements were performed with the NIST watt balance with the goal of obtaining an independent determination of the Planck constant. From the start, the experimenters were unaware of the precise value of the mass standard being used, i.e., the experiment was performed blindly to avoid experimenters' bias. The mass value was unveiled during a public talk and the corresponding value of the Planck constant was very close to the final result presented here. 

The value of the Planck constant is $h/\hn-1= 141(45)\times 10^{-9}$. The uncertainty budget is balanced. Five of the eight categories contribute a relative uncertainty of more than $\approx 15\times 10^{-9}$. Hence, it is difficult to reduce the uncertainty significantly, as all five categories would need to be improved.
 
During 2004-2010, the same experiment yielded lower values.  In 2010, the values shifted by $\approx 90\times 10^{-9}$. At the time of this writing, no satisfactory explanation for this increase has been found. The value after 2010 is in closer agreement to the value determined here.
 
\niii\ is a research apparatus designed to measure the Planck constant. For the new SI~\cite{Mills11}, a device is needed to realize mass based on a fixed value of $h$. A new watt balance, NIST-4 is currently under construction for this purpose. We hope to use \niii\ once more to compare the first results of NIST-4 to those from \niii.

An experiment of this scale was made possible through the help of our esteemed colleagues. First and foremost, we would like to thank our retired colleague, Edwin Williams. Ed was always available for ideas and provided gentle advice for solving many problems. Zeina Kubarych, Patrick Abbott, and Vincent Lee helped us by calibrating K85 and various other masses. We would like to thank them for their support in performing the blind measurement. Rand Elmquist and Marlin Kraft calibrated our resistors and shared their expertise in resistance metrology. The experts on Josephson Voltage systems, Sam Benz, Charlie Burroughs, Alain R\"ufnacht and Yi-hua Tang offered valuable advice on voltage metrology. We thank Jack Stone for calibrating our lasers. We would like to thank Jacques Liard for guidance on gravity measurements and corrections. Last, we would like to thank Bryan Waltrip for building us several programmable current sources.

\end{document}